\documentclass[12pt]{article}
\pdfoutput=1
\usepackage{epsfig,amsfonts,amsthm}
\usepackage[normalem]{ulem}
\usepackage{amsmath,amssymb}
\usepackage{array}
\usepackage{amsmath}
\usepackage{amsfonts}
\usepackage{amssymb}
\usepackage{subfig}
\usepackage{wrapfig}
\usepackage{graphicx}
\usepackage{slashed}
\usepackage{soul}
\newcommand{\be}{\begin{equation}}
\newcommand{\ee}{\end{equation}}
\newcommand{\bea}{\begin{eqnarray}}
\newcommand{\eea}{\end{eqnarray}}

\newcommand{\doublet}[2]{ \left( \begin{array}{c}#1 \\ #2 \end{array}\right) }

\usepackage{color}

\usepackage[normalem]{ulem}
   
\definecolor{grey}{cmyk}{0,0,0,0.75}
\definecolor{tangerine}{cmyk}{0,0.5,1,0}
\definecolor{darkgreen}{cmyk}{1,0,1,0.23} 
\definecolor{Red}{rgb}{1,0,0}
\definecolor{Blue}{rgb}{0,0,1}
\definecolor{Green}{rgb}{0,1,0}
\definecolor{Grey}{cmyk}{0,0,0,0.75}
\definecolor{Tangerine}{cmyk}{0,0.5,1,0}
\definecolor{Darkgreen}{cmyk}{1,0,1,0.23}
\definecolor{Cyan}{cmyk}{1,0,0,0}
\definecolor{Yellow}{cmyk}{0,0,1,0}

\def\lsim{\mathrel{\rlap{\lower4pt\hbox{\hskip1pt$\sim$}}
    \raise1pt\hbox{$<$}}}         
\def\gsim{\mathrel{\rlap{\lower4pt\hbox{\hskip1pt$\sim$}}
    \raise1pt\hbox{$>$}}}         

\def\beq{\begin{equation}}
\def\eeq{\end{equation}}
\def\bea{\begin{eqnarray}}
\def\eea{\end{eqnarray}}

\def\<{\left\langle}
\def\>{\right\rangle}

\newcommand{\GeV}{{\ensuremath\rm \,GeV}}

\usepackage[normalem]{ulem}

\def\lsim{\mathrel{\rlap{\lower4pt\hbox{\hskip1pt$\sim$}}
    \raise1pt\hbox{$<$}}}         
\def\gsim{\mathrel{\rlap{\lower4pt\hbox{\hskip1pt$\sim$}}
    \raise1pt\hbox{$>$}}}         

\def\beq{\begin{equation}}
\def\eeq{\end{equation}}
\def\bea{\begin{eqnarray}}
\def\eea{\end{eqnarray}}

\def\<{\left\langle}
\def\>{\right\rangle}

\newcommand{\gev}{\mathrm{\;GeV}} 
\newcommand{\bt}{\begin{tabular}}
\newcommand{\et}{\end{tabular}}

\usepackage{hyperref}

\begin{document}
\bibliographystyle{OurBibTeX}

\title{
\hfill ~\\[-30mm]
\begin{footnotesize}
\hspace{70mm}
HIP-2019-28/TH\\
\end{footnotesize}
\vspace{10mm}
\textbf{Dark CP-violation through the $Z$-portal
} }
\date{}

\author{\\[-5mm]
Venus  Keus
\\ \\
\emph{\small venus.keus@helsinki.fi}\\
\emph{\small Department of Physics and Helsinki Institute of Physics,}\\
\emph{\small Gustaf Hallstromin katu 2, FIN-00014 University of Helsinki, Finland}\\
\emph{\small School of Physics and Astronomy, University of Southampton,}\\
\emph{\small Southampton, SO17 1BJ, United Kingdom}\\[4mm]
  }

\maketitle

\vspace*{-7mm}

\begin{abstract}
\noindent
{Despite great agreement with experiment, the Standard Model (SM) of particle physics lacks a viable Dark Matter (DM) candidate and sufficient amount of CP-violation to account for the observed baryon excess in the universe. 
Non-minimal Higgs frameworks are economic extensions of the SM which could remedy these shortcomings.
Within the framework of a three Higgs doublet model, 
we introduce an extended dark sector which accommodates both DM and CP-violation. Such dark
sources of CP-violation do not contribute to the Electric Dipole Moments and are therefore unconstrained.
We present a novel mechanism in which the CP-violating dark particles only interact with the SM through the gauge bosons, primarily the $Z$ boson. Such $Z$-portal dark CP violation is realised in the regions of the parameter space where Higgs-mediated (co)annihilation processes are sub-dominant and have negligible contributions to the DM relic density. We show that such $Z$-portal CP violating DM can still thermalise and satisfy all experimental and observational bounds and discuss the implications of such phenomena for electroweak baryogenesis.
} 
 \end{abstract}

\thispagestyle{empty}
\vfill
\newpage
\setcounter{page}{1}

\section{Introduction}

The Standard Model (SM) of particle physics has been extensively tested and is in great agreement with experiment, with its last missing particle, the Higgs boson, discovered at the LHC in 2012 \cite{Aad:2012tfa,Chatrchyan:2012ufa}. 
No significant deviation from the SM has been detected at the LHC so far and the properties of the observed scalar are in agreement with those of the SM-Higgs boson \cite{Flechl:2019jnr,Aad:2019mbh}. 
However, SM falls
short of explaining several aspects of nature, such as the observed baryon asymmetry
in the universe and a viable candidate for Dark Matter (DM). 
Many astrophysical observations call for a DM particle which is stable on cosmological timescales, cold, non-baryonic, neutral and weakly interacting. A particle with such characteristics does not exist in the SM. 
Moreover, the amount of CP-violation provided by the SM is many orders of magnitude smaller than what is needed to generate the observed baryon excess \cite{Gavela:1993ts,Huet:1994jb,Gavela:1994dt}.

Therefore, it is widely accepted that one needs to consider beyond SM (BSM) scenarios in pursuit of the ultimate theory of nature.
The simplest BSM scenarios
which aim to conquer these shortcomings are non-minimal Higgs frameworks, suggesting that the observed scalar at the LHC is just one member of an extended scalar sector. 
The scalar potential is the least constrained sector in the SM and could provide new sources of CP-violation if extended. 
Also, non-minimal Higgs sectors with discrete symmetries could naturally accommodate Weakly Interacting Massive Particles (WIMPs) \cite{Jungman:1995df,Bertone:2004pz,Bergstrom:2000pn,Ivanov:2012hc}, the most well-studied DM candidates.
The stability of the WIMP is ensured by the conservation of the discrete symmetry yielding a relic abundance 
in agreement with the Planck experiment \cite{Ade:2015xua} through the freeze-out mechanism. 

Extensive studies have been carried out to an advanced level in one 
singlet or one doublet scalar extensions of the SM, namely the Higgs portal models and two Higgs doublet models (2HDMs) (see e.g. \cite{Englert:2011yb,Branco:2011iw} and references therein). 
These models, however, 
by construction can only partly provide a solution to the SM shortcomings.
Frameworks with a further extended scalar sector, such as three Higgs doublet models (3HDMs),
contain viable DM candidates, provide new sources of CP-violation and 
strongly first order phase transition as
the origin of the baryon asymmetry, 
contain inflaton candidates driving the inflation process in the beginning
of the universe and 
provide a solution to the fermion mass hierarchy problem, 
all in one framework owing to different symmetries and symmetry breaking patterns realisable in the scalar potential, with the symmetry breaking patterns determining the number of active (developing a vacuum expectation value (VEV)) and inert (without a VEV) multiplets.

The well-known $Z_2$ symmetric Higgs portal model \cite{Bertolami:2007wb} and the Inert Doublet Model (IDM) \cite{Deshpande:1977rw}, are in agreement with direct and indirect DM searches. However, they are severely constrained by LHC data from invisible Higgs branching ratio and Higgs signal strength bounds. 
In addition to that, the scalar potential in these models is inevitably CP-conserving.

If one were to abandon the idea of DM, the 2HDM scalar potential could, in general, accommodate CP-violation. However, 
such new sources of CP-violation modify the SM-Higgs couplings and 
contribute
to the Electric Dipole Moments (EDMs) of the neutron, electron, and certain atomic nuclei \cite{Chupp:2017rkp} and are, therefore, highly constrained by experiment \cite{Inoue:2014nva,Keus:2015hva,Keus:2017ioh,Yamanaka:2017mef}.
The purely scalar singlet extension of SM, however, is CP-conserving regardless of an apparent phase in the VEVs or the parameters of the potential. 
Since a scalar singlet does not have any gauge interactions, there is a large degree of arbitrariness in the definition of action of charge conjugation on the singlet which renders the model CP-conserving \cite{Branco:1999fs}.
One could summarise this discussion in the following list where $S$ represents a singlet and $\phi$ a doublet scalar. The superscripts ${Z_2}^+$ and ${Z_2}^-$ stand for even and odd $Z_2$ charges, respectively.
\begin{itemize}
\item
$
\left\{\begin{array}{c}
\phi^{{Z_2}^+},S^{{Z_2}^-} \\[2mm]
\phi_1^{{Z_2}^+},\phi_2^{{Z_2}^-}\\
\end{array}
\right.
$
with vacuum alignment $(v,0)$: very constrained DM, no CP-violation.
\item 
$\phi_1^{{Z_2}^+},\phi_2^{{Z_2}^-}$ with vacuum alignment $(v_1,v_2)$: very constrained CP-violation, no DM.
\end{itemize}
To accommodate CP-violation and a viable DM candidate one needs to go beyond the simple doublet or singlet extensions of the SM. 
If the active sector is extended to accommodate CP-violation and the solitary inert sector is to provide the DM candidate, one runs into the same limitations as in the 2HDM, the IDM and the Higgs portal model \cite{Grzadkowski:2009bt, Osland:2013sla}.
Extending the inert sector to accommodate both DM and new sources of CP-violation, on the other hand, relieves the model from these constraints.

Such dark/inert sources of CP-violation were introduced for the first time in \cite{Cordero-Cid:2016krd} and studied further in \cite{Keus:2016orl,Cordero:2017owj,Cordero-Cid:2018man,Cordero-Cid:2020yba} in a 3HDM framework where the CP-mixed dark/inert scalars interact with the SM particles through Higgs and the gauge bosons.
It was shown that dark CP-violation provides a handle on the otherwise fixed gauge-scalar couplings and opens up a large region of the parameter space to accommodate a CP-violating DM candidate in agreement with cosmological and collider experiments.
Moreover, since the inert sector is protected from directly coupling to the SM particles, the dark CP-violation receives no limitation from the EDM experiments.
Again, the following list aims to summarise this discussion.
\begin{itemize}
\item
$
\left\{\begin{array}{c}
\phi_1^{{Z_2}^+},\phi_2^{{Z_2}^+},S^{{Z_2}^-} \\[2mm]
\phi_1^{{Z_2}^+},\phi_2^{{Z_2}^+},\phi^{{Z_2}^-}\\
\end{array}
\right.
$
with 
$(v_1,v_2,0)$: very constrained DM, very constrained CP-violation.
\item 
$
\left\{\begin{array}{c}
\phi_1^{{Z_2}^+},\phi_2^{{Z_2}^-},\phi_3^{{Z_2}^-}  \\[2mm]
\phi_1^{{Z_2}^+},\phi_2^{{Z_2}^-},S^{{Z_2}^-}\\
\end{array}
\right.
$
with $(v,0,0)$: DM and CP-violation
\end{itemize}
It is important to note that when the extended inert sector contains a doublet and a singlet scalar, as opposed to two doublet scalars, the amount of CP-violation is inevitably reduced due to the presence of the singlet \cite{Azevedo:2018fmj}. In addition to that, there are fewer co-annihilation channels present for the DM candidate. As a result, the model is more constrained and less favourable to the 3HDM counterpart.

Here, we present a novel mechanism in which the CP-violating dark particles only interact with the SM through the gauge bosons, primarily the $Z$ boson. Such $Z$-portal dark CP violation is realised in the regions of the parameter space where
the Higgs portal couplings are of order $10^{-4}$ or less, and as a result the
Higgs-mediated (co)annihilation processes are sub-dominant and have negligible contributions to the DM relic density. We show that such $Z$-portal CP violating DM can still thermalise and satisfy all experimental and observational bounds and discuss the implications of such phenomena for electroweak baryogenesis.

The layout of the remainder of the paper is as follows. In section \ref{scalar-potential}, we present the details of the scalar potential and the theoretical and experimental limits on its parameters. In section \ref{selection}, we construct and justify our benchmark scenarios. In section \ref{Abundance}, we show the effect of dark CP-violation on the production and annihilation of DM and in section \ref{conclusion}, we conclude and present the outlook for our future studies.

\section{The extended scalar sector}
\label{scalar-potential}

A scalar potential extended by Higgs doublets, which is symmetric under a group $G$ of phase rotations, can be written as the sum of two parts: $V_0$ with terms symmetric under any phase rotation, and $V_G$ with terms symmetric under $G$ \cite{Ivanov:2011ae,Keus:2013hya}. As a result, a $Z_2$-symmetric 3HDM can be written as
\bea
\label{V0-3HDM}
V_{3HDM}&=&V_0+V_{Z_2}, \\
V_0 &=& - \mu^2_{1} (\phi_1^\dagger \phi_1) -\mu^2_2 (\phi_2^\dagger \phi_2) - \mu^2_3(\phi_3^\dagger \phi_3) \nonumber\\
&&+ \lambda_{11} (\phi_1^\dagger \phi_1)^2+ \lambda_{22} (\phi_2^\dagger \phi_2)^2  + \lambda_{33} (\phi_3^\dagger \phi_3)^2 \nonumber\\
&& + \lambda_{12}  (\phi_1^\dagger \phi_1)(\phi_2^\dagger \phi_2)
 + \lambda_{23}  (\phi_2^\dagger \phi_2)(\phi_3^\dagger \phi_3) + \lambda_{31} (\phi_3^\dagger \phi_3)(\phi_1^\dagger \phi_1) \nonumber\\
&& + \lambda'_{12} (\phi_1^\dagger \phi_2)(\phi_2^\dagger \phi_1) 
 + \lambda'_{23} (\phi_2^\dagger \phi_3)(\phi_3^\dagger \phi_2) + \lambda'_{31} (\phi_3^\dagger \phi_1)(\phi_1^\dagger \phi_3),  \nonumber\\
 V_{Z_2} &=& -\mu^2_{12}(\phi_1^\dagger\phi_2)+  \lambda_{1}(\phi_1^\dagger\phi_2)^2 + \lambda_2(\phi_2^\dagger\phi_3)^2 + \lambda_3(\phi_3^\dagger\phi_1)^2  + h.c., \nonumber
\eea
where the three Higgs doublets, $\phi_{1},\phi_2,\phi_3$, transform under the $Z_2$ group, respectively, as 
\be 
\label{generator}
g_{Z_2}=  \mathrm{\rm diag}\left(-1, -1, +1 \right). 
\ee

Note that adding the four $Z_2$-respecting terms in $V_{Z_2}$ ensures that our 3HDM potential is symmetric only under this $Z_2$ group \cite{Ivanov:2011ae} and no larger symmetry. There exist other $Z_2$-respecting terms
$ 
(\phi_1^\dagger\phi_2)(\phi_1^\dagger\phi_1), 
(\phi_1^\dagger\phi_2)(\phi_2^\dagger\phi_2),
(\phi_3^\dagger\phi_1)(\phi_2^\dagger\phi_3)
$
and
$ 
(\phi_1^\dagger\phi_2)(\phi_3^\dagger\phi_3), 
$
whose inclusion does not add to the phenomenology of the model\footnote{The terms $ 
(\phi_1^\dagger\phi_2)(\phi_1^\dagger\phi_1), \,(\phi_1^\dagger\phi_2)(\phi_2^\dagger\phi_2)$ only appear in inert scalar self-interactions and do not affect our tree-level analysis. Including the terms $(\phi_3^\dagger\phi_1)(\phi_2^\dagger\phi_3), \,(\phi_1^\dagger\phi_2)(\phi_3^\dagger\phi_3)$ resembles itself as a shift in the value of the $\lambda_{2,3}$ parameters. Our numerical analysis confirms that removing these terms does not lose the generality of the model.}. We have, therefore, set the coefficients of these terms to zero for simplicity.

The composition of the doublets is as follows:
\be 
\phi_1= \doublet{$\begin{scriptsize}$ H^+_1 $\end{scriptsize}$}{\frac{H_1+iA_1}{\sqrt{2}}},\quad 
\phi_2= \doublet{$\begin{scriptsize}$ H^+_2 $\end{scriptsize}$}{\frac{H_2+iA_2}{\sqrt{2}}}, \quad 
\phi_3= \doublet{$\begin{scriptsize}$ G^+ $\end{scriptsize}$}{\frac{v+h+iG^0}{\sqrt{2}}}, 
\label{explicit-fields}
\ee
where $\phi_1$ and $\phi_2$ are the $Z_2$-odd \textit{inert} doublets, $\langle \phi_1 \rangle = \langle \phi_2 \rangle =0$, and $\phi_3$ is the one $Z_2$-even \textit{active} doublet, $\langle \phi_3 \rangle =v/$\begin{scriptsize}$ \sqrt{2} $\end{scriptsize} $ \neq 0$, which plays the role of the SM Higgs doublet, with $h$ being the SM Higgs boson and $G^\pm,~ G^0$ the would-be Goldstone bosons. 
Note that the $Z_2$ charges assigned to each doublet are according to the $Z_2$ generator in Eq.(\ref{generator}): odd-$Z_2$ charge to the inert doublets, $\phi_1$ and $\phi_2$, and even-$Z_2$ charge to the active doublet, $\phi_3$. Therefore, the symmetry of the potential is respected by the vacuum $(0,0,v/$\begin{scriptsize}$ \sqrt{2} $\end{scriptsize}$)$. 

The CP-even scalar $h$ contained in the active doublet $\phi_3$ has the tree-level couplings of the SM Higgs boson. Thus CP-violation is only introduced in the \textit{inert} sector which is forbidden from mixing with the \textit{active} sector by the conservation of the $Z_2$ symmetry. As a result, the amount of CP-violation is not limited by EDMs. This phenomenon of unlimited dark CP-violation was first introduced in \cite{Cordero-Cid:2016krd}. The DM candidate, which is the lightest particle amongst the CP-mixed neutral fields from the inert doublets\footnote{We avoid regions of parameter space where one of the charged inert scalars is the lightest inert particle.}, is indeed stable due to the unbroken $Z_2$ symmetry. 

\subsection{Explicit CP-violation}
\label{section-2}
The parameters of the phase invariant part, $V_0$, are real by construction. We introduce explicit CP-violation through complex parameters, $\mu^2_{12}, \lambda_1,\lambda_2, \lambda_3$ in the potential in Eq.(\ref{V0-3HDM}).
Let us emphasize, however, that $\lambda_1$ (and other dark sector parameters $\lambda_{11},\lambda_{22},\lambda_{12}, \lambda'_{12}$) only concern inert scalars self-interactions and do not influence tree-level DM and collider phenomenology of the model. These parameters are only constrained through perturbative unitarity and positivity of $V$ bounds and do not play any role in our tree-level analysis. Their value has, therefore, been set to $0.1$. 

The parameters which are phenomenologically relevant are $\mu^2_3, \lambda_{33}$ which are fixed by the Higgs mass, and $\mu^2_{1},\mu^2_{2},\mu^2_{12}, \lambda_{31},\lambda_{23},\lambda'_{31},\lambda'_{23},\lambda_{2}, \lambda_{3}$ which appear in inert scalar masses and scalar couplings. The latter nine parameters are, in principle, independent. However, for simplicity, here we study the model in the \textit{dark democracy} limit \cite{Keus:2014jha,Keus:2015xya,Cordero-Cid:2016krd, Cordero-Cid:2018man}, where
\be 
\mu^2_1 =\mu^2_2 , \quad \lambda_3=\lambda_2 , \quad \lambda_{31}=\lambda_{23} ,\quad \lambda'_{31}=\lambda'_{23}.
\ee
The model is still explicitly CP-violating, provided $(\lambda_{22}- \lambda_{11} )
\left[\lambda_1 ({\mu^2_{12}}^*)^2-\lambda^*_{1}(\mu^2_{12})^2 \right] \neq 0$ \cite{Haber:2006ue,Haber:2015pua} after imposing the dark democracy limit\footnote{For a 3HDM to be CP-violating at least one CP-odd invariant has to be non-zero. Therefore, the non-vanishing CP-odd invariant condition provided here, is a sufficient condition but not necessary since any other non-zero CP-odd invariant ensures CP-violation in the model.}. 
Note that one can redefine the doublets as 
\be 
\left\{\begin{array}{c}
\phi_1 \to \phi_1 e^{i \theta_{12}/2} ~\\[2mm]
\phi_2 \to \phi_2 e^{-i \theta_{12}/2}\\[2mm]
\phi_3 \to \phi_3 ~~~~~~~~~
\end{array}
\right.
~~
\Longrightarrow 
~~
\left\{\begin{array}{c}
|\mu^2_{12}| e^{i \theta_{12}} \to |\mu^2_{12}|~~~~~~~ \\[2mm]
|\lambda_2| e^{i \theta_2} \to |\lambda_2| e^{i (\theta_2+\theta_{12})}
\end{array}
\right.
\ee
As a result, the only relevant CP-violating parameter\footnote{Recall that, as a dark parameter, $\lambda_1$ and its phase are not relevant for our tree-level analysis.} in the dark democracy limit
is $\theta_2+\theta_{12}$, the ``shifted'' phase of the $\lambda_2$ coupling,
which is referred to as $\theta_{\rm CPV}$ throughout the paper.

\subsection{The mass spectrum}
\label{minimization}
The point $(0,0,\frac{v}{\sqrt{2}})$ is the minimum of the potential when $ v^2= \frac{\mu^2_3}{\lambda_{33}}$.
The only active doublet, $\phi_3$, plays the role of the SM Higgs doublet, with $G^0,G^\pm$ as the massless Goldstone bosons, and $h$ as the SM-like Higgs with
$ m^2_{h}= 2\mu_3^2 =2\lambda_{33} v^2 = (125$ GeV$)^2$.

\subsubsection{The charged inert states}
The charged mass-squared matrix in the $(H^\pm_{1},H^\pm_{2})$ basis is calculated to be
\be 
\mathcal{M}^2_C= \left( \begin{array} {cc}
- \mu_{2 }^{2} + \frac{1}{2} \lambda_{23} v^{2} 
& -\mu_{12}^{2}  \\
- \mu_{12}^{2}   &  -\mu_{2}^{2} + \frac{1}{2} \lambda_{23} v^{2}  \end{array} \right) ,
\ee
with the two physical charged states, $S^\pm_{1,2}$, as eigenstates
\be 
S^\pm_{1}= \frac{H^\pm_{1} + H^\pm_{2}}{\sqrt{2}}, \qquad
S^\pm_{2}= \frac{H^\pm_{1} - H^\pm_{2}}{\sqrt{2}}.
\ee
The masses of the physical charged scalars, eigenvalues of the $\mathcal{M}^2_C$ matrix, are calculated to be 
\be 
m^2_{S^\pm_{1}}
=
 - \mu_2^2 - \mu_{12}^2 +\frac{1}{2}\lambda_{23}v^2, \qquad
m^2_{S^\pm_{2}}
=
- \mu_2^2 + \mu_{12}^2 + \frac{1}{2}\lambda_{23}v^2 ,
\ee
where we take $\mu_{12}^2 >0$ and, therefore, $m_{S_1^\pm} < m_{S_2^\pm}$. 

\subsubsection{The neutral inert states}
The neutral mass-squared matrix  in the $(H_1,H_2,A_1,A_2)$ basis is calculated to be
\be 
\mathcal{M}^2_N= \frac{1}{4}\left( \begin{array} {cccc}
\Lambda^+_c
& -2\mu^2_{12}
& -\Lambda_s
& 0 
\\[2mm]
-2\mu^2_{12}
& \Lambda^+_c
& 0 
& \Lambda_s 
\\[2mm]
-\Lambda_s
& 0 
&  \Lambda^-_c
& -2\mu^2_{12} 
\\[2mm]
0 
& \Lambda_s
& -2\mu^2_{12}
&  \Lambda^-_c 
\end{array} \right ),
\label{neutral-mass-squared}
\ee
where
\be  
\Lambda_s=  2\lambda_2 \, \sin \theta_{\rm CPV} \, v^2
\quad
\mbox{and}
\quad 
\Lambda^\pm_c =-2\mu^2_2 +
(\lambda_{23}+\lambda'_{23} \pm 2 \lambda _2  \cos \theta_{\rm CPV}) v^2,
\ee  
with the four CP-mixed neutral scalars, $S_{1,2,3,4}$, as eigenstates
\bea 
&& 
S_1 =\frac{\alpha H_1 -A_1+ \alpha H_2+A_2}{\sqrt{2} \; \sqrt{\alpha^2+1}},\qquad
S_2 =\frac{H_1 +\alpha A_1+ H_2  - \alpha A_2}{\sqrt{2} \; \sqrt{\alpha^2+1}},  \\[2mm]
&& 
S_3 =\frac{\beta H_1+A_1 -\beta H_2+A_2}{\sqrt{2} \; \sqrt{\beta^2+1}},
\qquad
S_4 =\frac{- H_1+\beta A_1  + H_2 +\beta A_2}{\sqrt{2}\; \sqrt{\beta^2+1}}, \nonumber
\eea
where $\alpha$ and $\beta$ are defined as
\be
\label{alpha-beta} 
\alpha 
= 
\frac{
- \mu^2_{12}
+v^2 | \lambda_2|  \cos\theta_{CPV}
-\Lambda^- }{v^2 | \lambda_2|  \sin\theta_{CPV}}
,\qquad
\beta = \frac{
-\mu^2_{12}
-v^2 | \lambda_2|  \cos\theta_{CPV}
+ \Lambda^+ }{v^2 | \lambda_2|  \sin\theta_{CPV}},
\ee
and $\Lambda^{\mp} $ as
\be  
\label{lambdas}
\Lambda^{\mp} =\sqrt{(\mu^2_{12})^2 + v^4|\lambda_2|^2 \mp 2 v^2 \mu^2_{12} |\lambda_2|\cos\theta_{CPV}} \; .
\ee
The masses of the physical neutral CP-mixed inert scalars, eigenvalues of the $\mathcal{M}^2_N$ matrix, are calculated to be
\bea
\label{masses-Ss}
m^2_{S_{1,2}} &=&  -\mu^2_2
+\frac{v^2}{2}(\lambda'_{23}+\lambda_{23}) \mp \Lambda^- ,
\\
m^2_{S_{3,4}} &=&  -\mu^2_2
+\frac{v^2}{2}(\lambda'_{23}+\lambda_{23}) \mp \Lambda^+  .
\nonumber
\eea
We take $S_1$ to be lightest inert particle and the DM candidate. Throughout the paper, the notations $S_1$ and DM will be used interchangeably.

\subsubsection{The span of $ \theta_{CPV}$}
To reproduce the results of \cite{Keus:2014jha,Keus:2015xya} with $\lambda_2 < 0$, we require\footnote{Note that for $\lambda_2 > 0$, the above augments hold provided the neutral inert particles are relabelled as $S_1 \leftrightarrow S_3$ and $S_2 \leftrightarrow S_4$.}
\be 
\frac{\pi}{2} < \theta_{CPV} < \frac{3\pi}{2} \qquad \Rightarrow \qquad m_{S_1} < m_{S_2}, m_{S_3},m_{S_4}.
\ee
Consequently, in the first and fourth quadrants $S_3$ becomes the lightest neutral inert particle,
\be
\left. \begin{array}{c}
0 < \theta_{CPV} < \frac{\pi}{2}\\[2mm]
\frac{3\pi}{2} < \theta_{CPV} < 2 \pi\\
\end{array}
\right\} 
\qquad \Rightarrow \qquad
m_{S_3} < m_{S_1}, m_{S_2},m_{S_4}.
\ee
At $\theta_{CPV} = \frac{\pi}{2}, \frac{3\pi}{2}$ where $\Lambda^+=\Lambda^-$, a mass degeneracy between neutral inert particles occurs,
\be
\theta_{CPV} = \frac{\pi}{2}, \frac{3\pi}{2}
\qquad \Rightarrow \qquad
\left\{ \begin{array}{c}
m_{S_1} = m_{S_3},\\[2mm]
m_{S_2} =m_{S_4}.\\
\end{array}
\right.
\label{s1s3-degeneracy}
\ee

The model is reduced to the CP-conserving limit when $\theta_{CPV} = 0,\pi$ which renders $S_{1,3}$ to CP-even and $S_{2,4}$ to CP-odd particles,
\be 
\theta_{CPV} = 0,\pi \qquad \Rightarrow \qquad
\mbox{CP-conserving limit}: \;
\left\{ \begin{array}{c}
S_{1,3} = \frac{H_1 \pm H_2}{\sqrt{2}},\\[2mm]
S_{2,4} =\frac{A_1 \pm A_2}{\sqrt{2}}.\\
\end{array}
\right. 
\label{CPC-limit}
\ee
We take all other parameters of the potential to be positive. 

\subsubsection{The input parameters of the model}
As independent input parameters of the model, we take
\be 
m_{S_{1}},\, m_{S_2}, \, m_{S^\pm_{1}}, \, m_{S^\pm_{2}}, \, \theta_{CPV}, \, g_{hDM}, 
\label{input-params}
\ee
where $g_{hDM} \equiv g_{S_1S_1h}$ is the Higgs-DM coupling, with the relevant terms in the Lagrangian appearing as 
\be 
\mathcal{L} \; \supset \; g_{ZS_iS_j} Z_\mu (S_i \partial^\mu S_j - S_j \partial^\mu S_i) +  \;
\frac{v}{2}g_{S_i S_i h} h S_i^2+ \;
v g_{S_i S_j h} h S_i S_j +  \;
v g_{S_i^\pm S_j^\mp h} h S_i^\pm S_j^\mp. \label{ghSS}
\ee
The parameters of the model can be written in terms of the physical observables in Eq.(\ref{input-params}):
\bea
\label{parameters}
&& \mu^2_{12} = \frac{1}{2}(m^2_{S^\pm_2} -m^2_{S^\pm_1}) ,
\\[1mm]
&&
\lambda_{23}=\frac{1}{v^2}
(2\mu^2_2+m^2_{S^\pm_2} +m^2_{S^\pm_1}),
\nonumber\\[1mm]
&&
\lambda'_{23}=\frac{1}{v^2}(m^2_{S_2}+m^2_{S_1}-m^2_{S^\pm_2}-m^2_{S^\pm_1}), \nonumber\\[1mm]
&&
|\lambda_2|= \frac{1}{v^2}
\left[\mu^2_{12}\cos\theta_{CPV}+ \frac{1}{4}\sqrt{(2\, \mu^2_{12} \cos\theta_{CPV})^2 + \left(m^2_{S_2}-m^2_{S_1}\right)^2-\left(m^2_{S^\pm_2}-m^2_{S^\pm_1}\right)^2} ~\right],
\nonumber\\[1mm]
&&
\mu^2_2= \frac{v^2}{2}g_{hDM} -\frac{v^2}{\alpha^2+1}
\left(
2 \,\alpha \, \sin\theta_{CPV}  +(\alpha^2-1)\cos\theta_{CPV}
\right)|\lambda_2|
-\frac{1}{2}(m^2_{S_2}+
m^2_{S_1}).  \nonumber
\eea
Using these relations, all other relevant masses and couplings can be expressed in terms of the independent input parameters of the model, for example, the masses of other inert scalars:
\be 
m^2_{S_{3,4}} =m^2_{S_{1}} + \Lambda^- \mp \Lambda^+, \qquad \mbox{where} \quad (\Lambda^+)^2 = (\Lambda^-)^2 +4 v^2 \mu^2_{12} |\lambda_2|\cos\theta_{CPV}.
\ee
The neutral scalar-gauge couplings are derived to be
\bea 
&&
|g_{Z S_1 S_3}| = |g_{Z S_2 S_4}| =  g^{CPC}_{Z}(\frac{\alpha+\beta}{\sqrt{\alpha^2+1}\sqrt{\beta^2+1}}) , \\
&&
|g_{Z S_1 S_4}| =| g_{Z S_2 S_3}| =  g^{CPC}_{Z} (\frac{\alpha\beta-1}{\sqrt{\alpha^2+1}\sqrt{\beta^2+1}}),
\nonumber
\eea
where $g^{CPC}_{Z}=\frac{e}{2 \,c_{\theta_W} s_{\theta_W}}$ is the $|g_{ZS_iS_j}|$ value in the CP-conserving limit, with $e$ the elementary charge and $c_{\theta_W}, s_{\theta_W}$ the sine and cosine of the weak mixing angle. Note that 
\bea 
&& g^2_{Z S_{1} S_3} + g_{Z S_{1} S_4}^2 = (g^{CPC}_{Z})^2, 
\qquad 
g^2_{Z S_{2} S_3} + g_{Z S_{2} S_4}^2 =  (g^{CPC}_{Z})^2
\eea
where $g_{ZS_1S_2}=g_{ZS_3S_4}=0$ in the dark democracy limit.
Similarly, the charged scalar-gauge couplings are derived to be
\bea
&&|g_{W^\pm S_{1}^\mp S_1}| = |g_{W^\pm S^\mp_{2}S_2}| = g^{CPC}_{W} (\frac{\alpha}{\sqrt{\alpha^2+1}}) ,
\qquad
|g_{W^\pm S^\mp_{1}S_2}| = |g_{W^\pm S^\mp_{2}S_1}| =g^{CPC}_{W} (\frac{1}{\sqrt{\alpha^2+1}}) \nonumber\\[1mm]
&& 
|g_{W^\pm S_{1}^\mp S_3}| = |g_{W^\pm S^\mp_{2}S_4}| = g^{CPC}_{W} (\frac{1}{\sqrt{\beta^2+1}}) ,
\qquad
|g_{W^\pm S^\mp_{1}S_4}| = |g_{W^\pm S^\mp_{2}S_3}| =g^{CPC}_{W} (\frac{\beta}{\sqrt{\beta^2+1}}) \nonumber
\eea
where $g^{CPC}_{W}=\frac{e}{s_{\theta_W}}$ is the $|g_{W^\pm S_{i}^\mp S_j}| $ value in the CP-conserving limit. 
Note that, unlike the CP-conserving limit, the strength of gauge-scalar interactions depend on the parameters $\alpha$ and $\beta$ in Eq.(\ref{alpha-beta}), which in turn depend on $m_{S_i}$ in the presence of CP-violation.

\subsection{Constraints on the parameter space}
\label{constraints}

The parameter space of the model is constrained by theoretical, observational and experimental bounds which are satisfied in all our benchmark scenarios to follow:

\begin{enumerate}
\item 
To satisfy theoretical constraints, we require the potential to be bounded from below and for the Hessian to be positive-definite \cite{Cordero-Cid:2016krd}, using the conservative sufficient limits of
\be
\lambda_{ii} > 0, \quad 
\lambda_{ij} + \lambda'_{ij} > -2 \sqrt{\lambda_{ii}\lambda_{jj}}, \quad 
|\lambda_{1,2,3}|< |\lambda_{ii}|, |\lambda_{ij}|, |\lambda'_{ij}| , \quad i\neq j = 1,2,3.
\ee
We take all couplings to be $|\lambda_i| \leq\,4\,\pi$ in accordance with perturbative unitarity limits. 

\item 
Parameterised by the EW oblique parameters $S,T,U$ \cite{Altarelli:1990zd}-\cite{Maksymyk:1993zm}, inert particles $S_i, S_i^{\pm}$ may introduce important radiative corrections to gauge boson propagators.
We impose a $2\sigma$ agreement with EW Precision Observables (EWPOs) at $95 \%$ Confidence Level (CL) \cite{Baak:2014ora},
\be 
S = 0.05\pm0.11, \quad T = 0.09\pm0.13, \quad U = 0.01\pm0.11.
\ee
Similar to the 2HDM, this condition requires each charged state to be close in mass with a neutral state, in the dark sector \cite{Dolle:2009fn}.

\item 
The contribution of the inert scalars to the total decay width of the EW gauge bosons  constrains the masses of the inert scalars to be \cite{Agashe:2014kda}
\be 
\label{eq:gwgz}
m_{S_i}+m_{S_{1,2}^\pm}\,\geq\,m_{W^\pm}, \quad
\,m_{S_i}+m_{S_j}\,\geq\,m_Z, \quad
\,2\,m_{S_{1,2}^\pm}\,\geq\,m_Z, \quad
i,j=1,2,3,4.
\ee

\item  
Non-observation of charged scalars puts a model-independant lower bound on their mass \cite{Lundstrom:2008ai,Cao:2007rm,Pierce:2007ut} and an upper bound on their lifetime \cite{Heisig:2018kfq} to be
\be 
m_{S_{1,2}^\pm}\,\geq\,70\,\GeV, \qquad
\tau_{S_{1,2}^\pm}\,\leq\,10^{-7} \, s \; \Rightarrow \;
\Gamma^\text{tot}_{S_{1,2}^\pm}\,\geq\,6.58\,\times\,10^{-18}\,\GeV,
\ee
to guarantee their decay within the detector. 
In all our benchmark scenarios, the mass of both charged scalars is above 95 GeV and their decay width, primarily to $S^{\pm}_i \to S_j W^\pm$, is of the order of $10^{-1}$ GeV, which is well within limits.

\item
Any model introducing new decay channels for the SM-Higgs boson is constrained by an upper limit on the Higgs total decay width, $\Gamma^h_\text{tot}\,\leq\,9$ MeV \cite{CMS:2018bwq}, and Higgs signal strengths \cite{Khachatryan:2016vau,Aaboud:2018xdt,Sirunyan:2018ouh}. 
In our model, the SM-like Higgs could decay to a pair of inert scalars, provided $m_{S_i}+m_{S_j} < m_h$ and $S_{i,j}$ are long-lived enough ($\tau\,\geq\,10^{-7}$ s). 
As a result, $S_{i,j}$  will not decay inside the detector and therefore contribute to the Higgs \textit{invisible} decay, $h \to S_i S_j$, with a branching ratio of
\be
\textrm{BR}(h \to S_i S_j) = \frac{\sum_{i,j} \Gamma(h\to S_iS_j)}{\Gamma_h^{\rm SM}+\sum_{i,j} \Gamma(h \to S_iS_j)}, 
\label{inv_all}
\ee
with
\be
\Gamma(h\to S_i S_j)=\frac{g_{h S_i S_j}^{2}v^2}{32\pi m_{h}^3}
\biggl( \left(m_h^2-(m_{S_i}+m_{S_j})^2 \right) \left(m_h^2-(m_{S_i}-m_{S_j})^2 \right)\biggr)^{1/2},
\ee
which sets strong limits on the Higgs-inert couplings. 
Moreover, the partial decay $\Gamma(h\to \gamma\gamma)$ receives contributions from the inert charged scalars. 
The combined ATLAS and CMS Run I results for Higgs to $\gamma\gamma$ signal strength require $\mu_{\gamma \gamma} = 1.14^{+0.38}_{-0.36}$ \cite{Khachatryan:2016vau}. 
In Run II, ATLAS reports $\mu_{\gamma \gamma} = 0.99^{+0.14}_{-0.14}$ \cite{Aaboud:2018xdt}, and CMS reports $\mu_{\gamma \gamma} = 1.18^{+0.17}_{-0.14}$ \cite{Sirunyan:2018ouh} with both of which we are in  $2\sigma$  agreement.

\item
Reinterpretion of LEP 2 and LHC Run I searches for Supersymmetric (SUSY) particles (mainly sneutrinos and sleptons) for the IDM  \cite{Lundstrom:2008ai,Belanger:2015kga} excludes the region of parameter space where the following conditions are simultaneously satisfied ($i=2, ... ,4$):
\be 
\label{eq:leprec}
m_{S_1}\,\leq\,80\,\gev,\,\, ~~
m_{S_i}\,\leq\,100\,\gev,\,~~
\Delta m {(S_1,S_i)}\,\geq\,8\,\gev.
\ee
We take these limits into account for our DM candidate paired with any other neutral scalar.
We also check the validity of our benchmark scenarios against LHC searches for new particles in accordance with the analysis for the IDM \cite{Kalinowski:2018ylg}.

\item 
DM relic density measurements from the Planck experiment \cite{Ade:2015xua},
\be
\label{eq:planck}
\Omega_{DM}\,h^2\,=\,0.1197\,\pm\,0.0022,
\ee
require the relic abundance of the DM candidate to lie within these bounds if it constitutes 100\% of DM  in the universe.

A DM candidate with $\Omega_{\rm DM} h^2 $ smaller than the observed value is allowed; however, an additional DM candidate is needed to complement the missing relic density. Regions of the parameter space corresponding to values of $\Omega_{\rm DM}h^2$ larger than the Planck upper limit are excluded.

We impose a $3\sigma$ agreement with the observation on the relic abundance of our DM candidate, $S_1$.

\item 
The latest XENON1T results for DM direct detection experiments \cite{Aprile:2018dbl} and FermiLAT results for indirect detection searches \cite{Fermi-LAT:2016uux} do not constrain the model any further. Having set the Higgs portal couplings to zero in our benchmark scenarios, the largest direct detection cross section is $\sigma_{DM-N} \approx 10^{-14}\; pb$ and the largest indirect detection cross section is $\langle v\sigma \rangle \approx 10^{-32} ~ cm^3/s$, both of which are well below the limits \cite{Billard:2013qya}.

\end{enumerate}

\section{DM abundance and the selection of benchmarks}
\label{selection}

The relic abundance of the DM candidate, $S_1$, after freeze-out is given by the solution of the Boltzmann equation,
\be 
\frac{d n_{S_1}}{dt} = 
- 3\, H \,n_{S_1} - \langle \sigma_{eff}\, v \rangle \, \left[ (n_{S_1})^2 - (n^{eq}_{S_1})^{2} \right],
\ee
where $n_{S_1}$ is the number density of the $S_1$ particle, $H$ is the Hubble parameter, and $n^{eq}_{S_1}$ is the number density of $S_1$ at equilibrium. 
The thermally averaged effective (co)annihilation cross section, $\langle \sigma_{eff}\, v \rangle$, receives contributions from all relevant annihilation processes of any $S_i S_j$ pair into SM particles, so that
\be 
\langle \sigma_{eff} v \rangle = 
\sum_{i,j} \langle \sigma_{ij}\, v_{ij} \rangle \,\frac{n^{eq}_{S_i}}{n^{eq}_{S_1}} \, \frac{n^{eq}_{S_j}}{n^{eq}_{S_1}},
\qquad
\mbox{where}
\qquad
\frac{n^{eq}_{S_i}}{n^{eq}_{S_1}} \sim \exp({-\frac{m_{S_i} - m_{S_1}}{T}}).
\ee
However, only processes with the  ${S_i} - {S_1}$ mass splitting comparable to the thermal bath temperature $T$ provide a sizeable contribution.

A common feature of non-minimal Higgs DM models is that in a large region of the parameter space the most important process for DM annihilation is through the
$S_1 S_1 \to h_{\rm SM} \to f \bar f$
channel whose efficiency depends on both the DM mass and the Higgs-DM coupling. 
In the region where $m_{\rm DM} < m_h/2$, generally one requires a large Higgs-DM coupling in order to produce relic density in agreement with Eq.(\ref{eq:planck}).
However, such large Higgs-DM coupling leads to large direct detection and indirect detection cross sections and significant deviations from SM-Higgs coupling measurements, which are ruled out by experimental and observational data. 
On the other hand, a small Higgs-DM coupling fails to annihilate the DM candidate effectively and leads to the over-closure of the universe.
This is where co-annihilation processes play an important role as they can contribute to changes in the DM relic density.
 
In models with extended dark sectors, in addition to the standard Higgs mediated annihilation channels of DM, there exists the possibility of co-annihilation with heavier states, provided they are close in mass \cite{Cordero-Cid:2016krd,Cordero:2017owj,Cordero-Cid:2018man,Keus:2014jha,Keus:2015xya}. The relevance of this effect depends not only on the DM mass and the mass splittings but also on the strength of the standard DM annihilation channel. 

It is worth pointing out that in the IDM, where by construction CP-violation is not allowed, the only co-annihilation process is through the $Z$-mediated $H\,A \to Z \to f \bar f$ channel whose sub-dominant effect fails to revive the model in the low mass region. 
Extending the inert sector,
as shown in \cite{Cordero:2017owj,Keus:2014jha,Keus:2015xya} in the CP-conserving limit, opens up several co-annihilation channels, both Higgs-mediated $H_1\,H_2 \to h \to f \bar f$ and $Z$-mediated $H_1\,A_{1,2} \to Z \to f \bar f$. However, their collective contribution to DM co-annihilation is not sufficient and one still needs a non-zero Higgs-DM coupling to satisfy relic density bounds.
Introducing CP-violation in the extended dark sector \cite{Cordero-Cid:2016krd,Cordero-Cid:2018man,Fuyuto:2019vfe} opens up many co-annihilation channels through the Higgs and $Z$ bosons, $S_i\,S_j \to h/Z \to f \bar f$, which can significantly affect the DM phenomenology.
In fact, the $Z$-mediated co-annihilations can be strong enough to relieve the model of the need for any Higgs-mediated (co)annihilation processes. 

To show the effect of $Z$ portal CP-violation on the abundance of DM, we set the Higgs-DM coupling to zero, $g_{hDM}=0$, thereby removing the main DM annihilation process, $S_1 S_1 \to h \to f \bar f$. All other $S_iS_j h$ vertex coefficients are also reduced to a point where their resulting co-annihilation processes have negligible contributions to the DM relic density. So, the only communication between the dark sector and the visible sector is through the gauge bosons $W^\pm$ and $Z$ .

It is important to note that the phenomenon of dark CP-violation is not realisable in purely scalar singlet extensions of the SM. 
An extended dark sector with a doublet plus a singlet could accommodate dark CP-violation; however, the presence of the singlet dilutes the CP-violating effects, since it has no couplings to SM gauge bosons. We would like to point out that this is the reason Ref. \cite{Azevedo:2018fmj} does not find points leading to 100\% of DM relic density in their low mass region. Furthermore, the effects of dark CP-violation which is through the $ZZZ$ observable \cite{Grzadkowski:2016lpv}, in their model is considerably smaller \cite{Cordero-Cid:2018man,Cordero-Cid:2020yba}.

In the region of the parameter space where Higgs portal interactions are negligible ($g_{hDM} \approx 10^{-4}$), the total DM annihilation cross section receives contributions from the following:
\begin{itemize}
\item 
\textbf{DM annihilation processes:}
\be 
S_1 S_1 \to V V , \qquad
S_1 S_1 \to V V^* \to V f f', \qquad 
S_1 S_1 \to V^* V^* \to f f' f f' \,,
\label{annihilation-1}
\ee
where $V$ is any of the SM gauge bosons. In the $m_{DM} < m_{W^\pm}$ region, the processes with off-shell gauge bosons dominate over the ones with on-shell gauge bosons.

\item 
\textbf{DM co-annihilation processes:}
\be 
S_1 S_{2,3,4} \to Z^* \to f \bar f, \qquad S_1 S^\pm_{1,2} \to W^{\pm *} \to f f' \, ,
\label{annihilation-2}
\ee
where the co-annihilating dark scalars are up to 20\% heavier than the DM candidate.

\item 
\textbf{(co)annihilation of other dark states:}
\be  
S_i S_i \to V V , \quad
S_i S_i \to V V^* \to V f f', \quad 
S_i S_i \to V^* V^* \to f f' f f', \quad 
S_i S_j \to V^* \to f f', 
\label{annihilation-3}
\ee
where $S_i\neq S_j$ are any of the dark scalars $ S_{2,3,4}\,, S^\pm_{1,2}$ which are all close in mass.


\end{itemize}

Taking all such processes into account, we define the following benchmark scenarios with distinct DM phenomenology.
It is convenient to introduce the mass splittings between the DM candidate and other inert scalars as
\be
\delta_{12} = m_{S_2} - m_{S_1}, \qquad \delta_{c} = m_{S_2^\pm} - m_{S_1^\pm} , \qquad \delta_{1c} = m_{S_1^\pm} - m_{S_1}\, .
\ee

\subsubsection*{Benchmarks of type 1:}
In agreement with reinterpreted SUSY searches in Eq.(\ref{eq:leprec}), we devise two benchmark scenarios in the low mass region, $45 \gev <m_{S_1} \leq 80 \gev$, 
\bea
&&\mbox{\textbf{B$_1$D$_{4}$C$_{1}$}}: \; \delta_{12} =4 \GeV,\quad \delta_{c} = 1\GeV,\quad \delta_{1c} = 50\GeV ,  \nonumber \\
&&\mbox{\textbf{B$_1$D$_{8}$C$_{1}$}}: \; \delta_{12} =8 \GeV,\quad \delta_{c} = 1\GeV,\quad \delta_{1c} = 50\GeV , \nonumber  
\eea
where all neutral inert particles are close in mass and are much lighter than the inert charged particles, 
\be
m_{S_1} \sim m_{S_3}\sim m_{S_2} \sim m_{S_4} \ll m_{S_1^\pm} \sim m_{S_2^\pm} \,.
\ee 
If one were to ignore these reinterpreted SUSY bounds, one can also construct benchmark scenarios with larger mass splittings,
\bea
&& \mbox{\textbf{B$_1$D$_{12}$C$_1$}}: \; \delta_{12} =12 \GeV, \quad\delta_{c} = 1\GeV, \quad\delta_{1c} = 50\GeV,  
\nonumber  \\
&& \mbox{\textbf{B$_1$D$_{20}$C$_1$}}: \; \delta_{12} =20 \GeV, \quad\delta_{c} = 1\GeV, \quad\delta_{1c} = 50\GeV . 
\nonumber 
\eea
With a larger $\delta_{12}$, the neutral inert particles split into two groups, with $S_1$ and $S_3$ close in mass and lighter than $S_2$ and $S_4$ which are also close in mass, and all lighter than the charged inert scalars,
\be 
m_{S_1} \sim m_{S_3} \; \lesssim \;  m_{S_2} \sim m_{S_4} \;\ll \; m_{S_1^\pm} \sim m_{S_2^\pm}  \,.
\ee

\subsubsection*{Benchmarks of type 2:}
In the low mass region, $45 \gev <m_{S_1} \leq 80 \gev$, in agreement with reinterpreted SUSY searches in Eq.(\ref{eq:leprec}), we define 
\bea
&&\mbox{\textbf{B$_2$D$_{55}$C$_{1}$}}: \; \delta_{12} =55 \GeV, \quad \delta_{c} = 1\GeV, \quad \delta_{1c} = 50\GeV, \nonumber  \\
&&\mbox{\textbf{B$_2$D$_{55}$C$_{15}$}}: \; \delta_{12} =55 \GeV, \quad \delta_{c} = 15\GeV, \quad \delta_{1c} = 50\GeV,  \nonumber 
\eea
where only one neutral inert particle is close in mass with the DM candidate, 
\be
m_{S_1} \sim m_{S_3} \ll m_{S_2} \sim m_{S_4} \sim  m_{S_1^\pm} \sim m_{S_2^\pm} \, .
\ee

\subsubsection*{Benchmarks of type 3:}
In the heavy mass region $m_{S_1} \geq 80 \gev$, the reinterpreted SUSY bounds in Eq.(\ref{eq:leprec}) do not apply any more, so any $\delta_{12}$ mass splitting is allowed. Moreover, the charged particles are allowed to be close in mass with the DM. We define three benchmark scenarios:
\bea
\mbox{\textbf{B$_3$D$_{5}$C$_1$}}: &&
\; \delta_{12} =5 \GeV,\quad \delta_{c} = 1\GeV, \quad \delta_{1c} = 1\GeV, \nonumber
\\
\mbox{where} &&
 m_{S_1} \sim m_{S_3} \sim m_{S_2} \sim m_{S_4} \sim  m_{S_1^\pm} \sim m_{S_2^\pm} \,, 
\\
\mbox{\textbf{B$_3$D$_{55}$C$_1$}}: &&
\; \delta_{12} =55 \GeV, \quad\delta_{c} = 1\GeV, \quad\delta_{1c} = 1\GeV , \nonumber \\
\mbox{where} &&
 m_{S_1} \sim m_{S_3} \sim  m_{S_1^\pm} \sim m_{S_2^\pm} \ll m_{S_2} \sim m_{S_4} \,,
\\
\mbox{\textbf{B$_3$D$_{55}$C$_{22}$}}: &&
\; \delta_{12} =55 \GeV, \quad\delta_{c} = 22 \GeV, \quad\delta_{1c} = 1 \GeV ,  \nonumber\\
\mbox{where} &&
 m_{S_1} \sim m_{S_3} \sim  m_{S_1^\pm}  \ll  m_{S_2^\pm} \sim m_{S_2} \sim m_{S_4} \,.
\eea

\section{The effect of dark CP-violation on the abundance}
\label{Abundance}

\subsection{Benchmarks of type 1}
\begin{figure}[t]
\begin{center}
\includegraphics[scale=0.55]{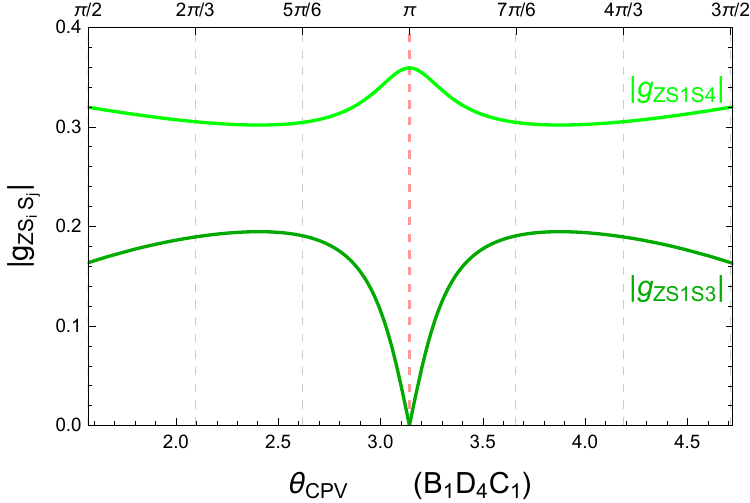}
\hspace{-2mm}
\includegraphics[scale=0.55]{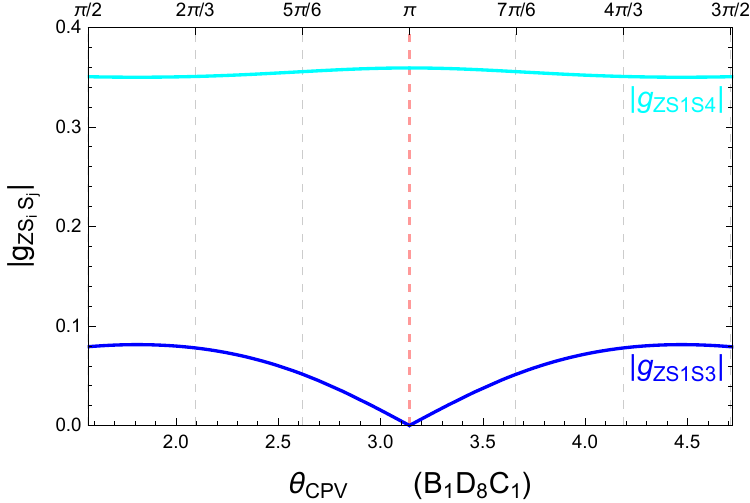}\\[2mm]
\includegraphics[scale=0.55]{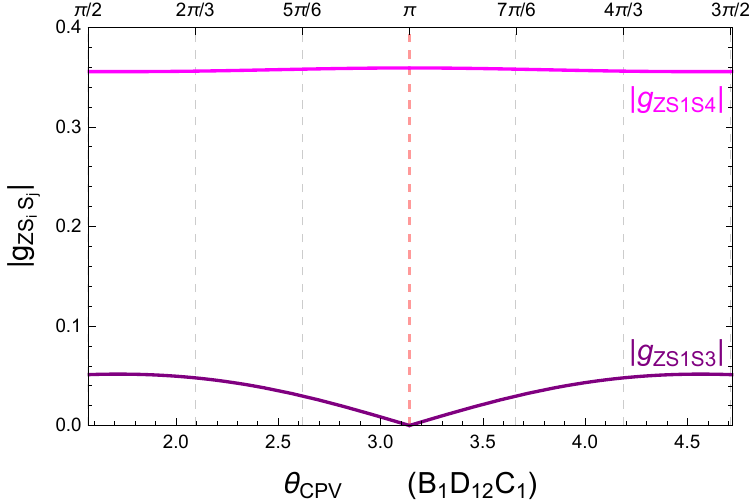}
\hspace{-2mm}
\includegraphics[scale=0.55]{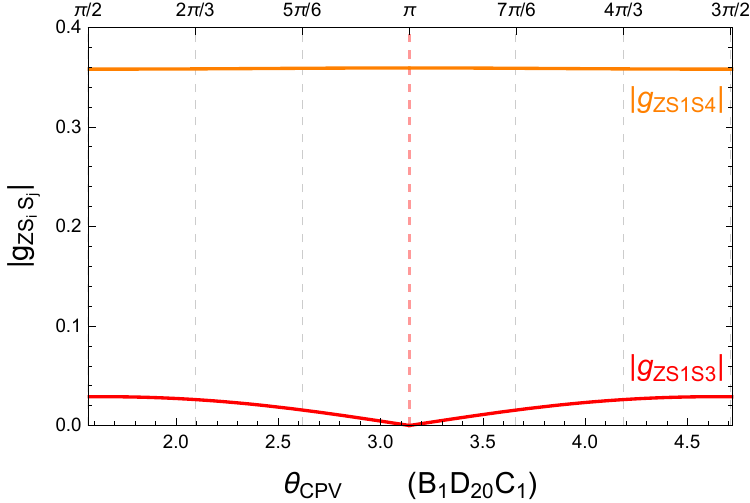}
\caption{The relevant and non-negligible $ZS_iS_j$ couplings in type 1 scenarios for an exemplary $m_{S_1}$ of 57 GeV.}
\label{B1-gZSiSj-fig}
\end{center}
\end{figure}

In benchmark scenarios of type 1, all neutral inert particles are relatively close in mass and are much lighter than the inert charged particles.
Here the main co-annihilation channel is through the
$S_iS_j \to Z^* \to f \bar f$ processes. 
With very small $S_1-S_2$ mass splitting, in B$_1$D$_{4}$C$_{1}$ and B$_1$D$_{8}$C$_{1}$ scenarios, the co-annihilation of DM with other neutral scalars is so strong that DM is under-produced irrespective of the CP-violating angle.
A larger $S_1-S_2$ mass splitting, as in B$_1$D$_{12}$C$_{1}$ and B$_1$D$_{20}$C$_{1}$ scenarios, weakens these co-annihilation processes and leads to a larger relic abundance of $S_1$.
The efficiency of the co-annihilation process also depends on the strength of the $ZS_iS_j$ coupling.
Figure \ref{B1-gZSiSj-fig} shows the strength of the relevant and non-negligible $ZS_iS_j$ couplings in all four scenarios for an exemplary $m_{S_1}$ of 57 GeV. 
As expected the $g_{ZS_1S_3}$ coupling vanishes at $\theta_{CPV}=\pi$ i.e. in the CP-conserving limit which is where $S_1$ and $S_3$ are reduced to two CP-even particles as discussed in Eq.(\ref{CPC-limit}). 

\begin{figure}[hhh]
\begin{center}
\includegraphics[scale=0.54]{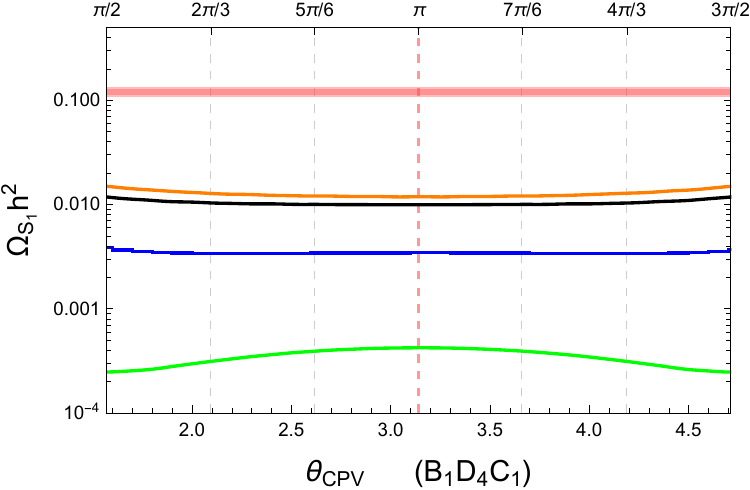}
\hspace{-3mm}
\includegraphics[scale=0.54]{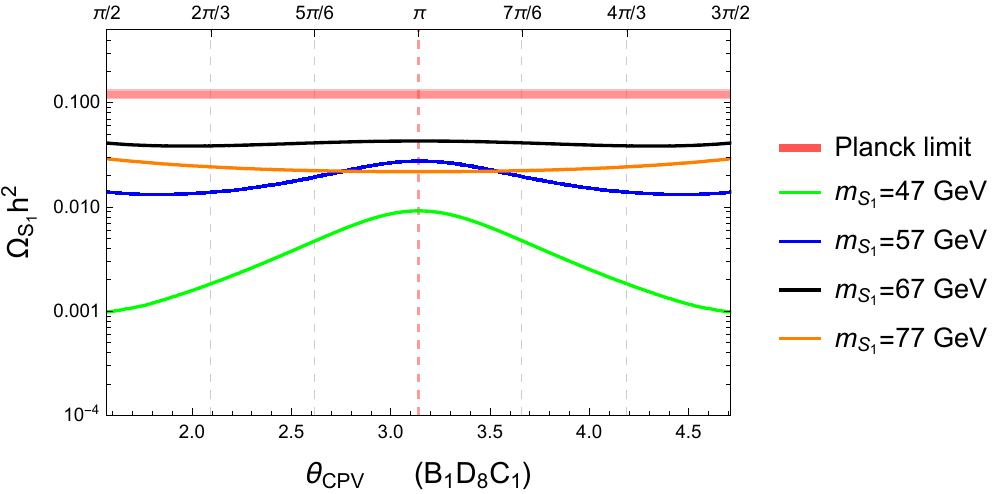}\\[2mm]
\includegraphics[scale=0.54]{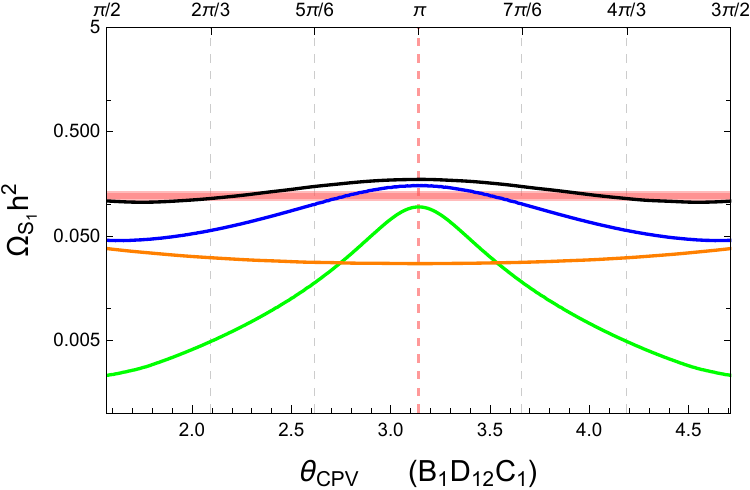}
\hspace{-3mm}
\includegraphics[scale=0.54]{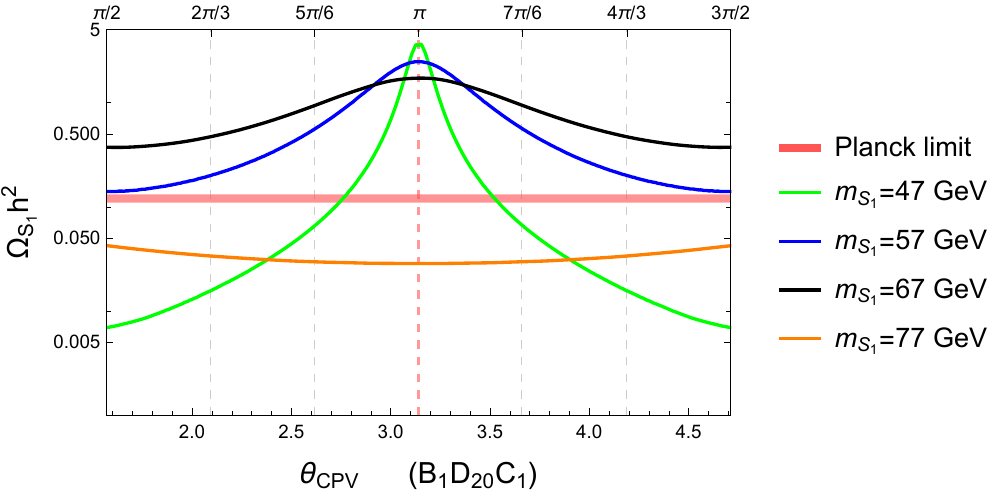}
\caption{The relic abundance of $S_1$ for different DM masses in type 1 benchmark scenarios. The horizontal red band shows the Planck observation limit on the abundance of DM.}
\label{B1-RelicAngle-fig}
\end{center}
\end{figure}

Note that as $\delta_{12}$ increases from B$_1$D$_4$C$_1$ to B$_1$D$_{20}$C$_1$ scenarios which reduces the co-annihilation probability of $S_1$ with other neutral dark particles, also the coupling of the main co-annihilation channel, $g_{ZS_1S_3}$ is reduced. As a result, the DM abundance is considerably larger in the latter scenarios.
Figure \ref{B1-RelicAngle-fig} shows the abundance of $S_1$ for different DM masses in all type 1 scenarios. 

\begin{figure}[h!]
\begin{center}
\includegraphics[scale=0.55]{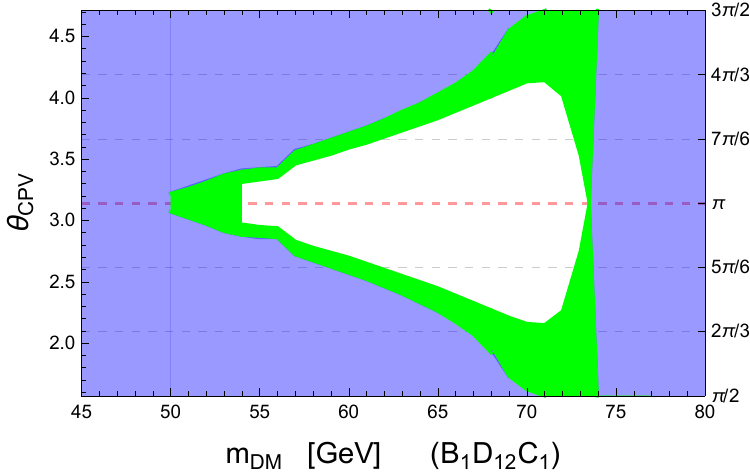}
\hspace{-4mm}
\includegraphics[scale=0.55]{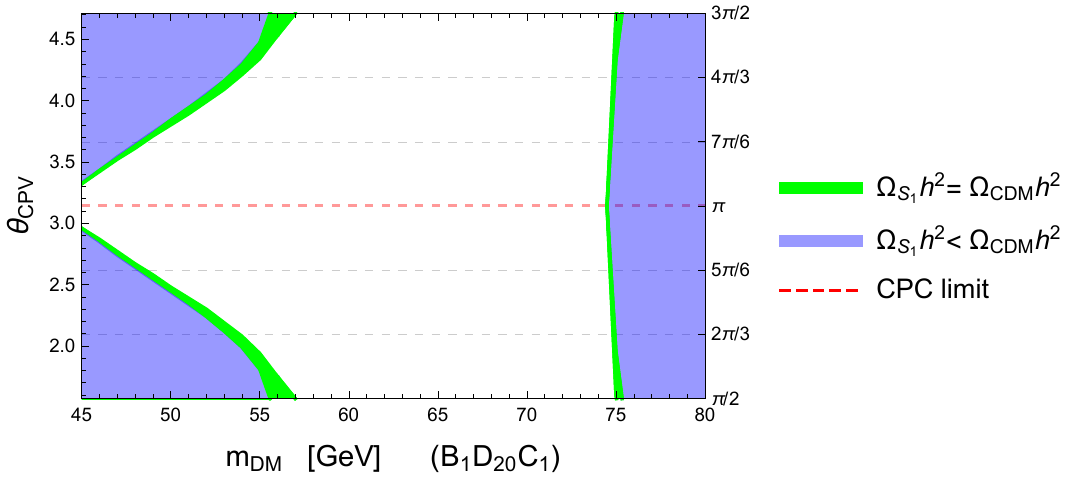}
\caption{Regions producing 100\% of DM in green and under-producing DM in blue in the $\theta_{CPV}$-$m_{DM}$ plane. The horizontal dashed red line represents the CP-conserving limit.}
\label{B4D12-B4D20-AngleMass-fig}
\end{center}
\end{figure}

It is worth noting that in a given benchmark scenario as DM mass increases to values comparable with $m_{W^\pm}$ and $m_{Z}$, the $S_iS_i \to VV$ annihilation channels open up and reduce the DM number density. As a result, DM is always under-produced for $m_{S_1} \gtrsim 80$ GeV. Figure \ref{B4D12-B4D20-AngleMass-fig} shows the regions with correct abundance and under-abundance of DM in B$_1$D$_{12}$C$_{1}$ and B$_1$D$_{20}$C$_{1}$ scenarios.
At $\theta_{CPV}=\pi$, where the model is CP-conserving, the strength of the $S_iS_jZ$ gauge couplings is fixed. As a result, the intermediate mass region $54 \gev \lesssim m_{S_1} \lesssim 70 \gev $ is ruled out due to the over production of DM.
Varying the CP-violating phase, $\theta_{CPV}$, changes the strength of the $S_iS_jZ$ couplings and the $S_1$-$S_3$ mass splitting, with the smallest mass splitting close to the $\pi/2$ and $3\pi/2$ boundaries. Therefore, one expects a more effective co-annihilation of $S_1$ with $S_3$ and a smaller relic density as $\theta_{CPV}$ moves away from the CP-conserving limit and towards the maximum CP-violation at the $\pi/2$ and $3\pi/2$ boundaries.
Figure \ref{B4D12-B4D20-AngleMass-fig} illustrates this behaviour in B$_1$D$_{12}$C$_{1}$ and B$_1$D$_{20}$C$_{1}$ scenarios. Note that in the latter scenario, due to the large $\delta_{12}$, the intermediate mass region over-produces DM regardless of the size of the $S_1$-$S_3$ mass splitting.

\subsection{Benchmarks of type 2}
In both 
B$_2$D$_{55}$C$_1$ and B$_2$D$_{55}$C$_{15}$ 
scenarios, with only $S_3$ close in mass with the DM, there exists only one co-annihilation channel, namely the 
$S_1S_3 \to Z^* \to f \bar f$ channel, which dictates the behaviour of the model in the low mass region.
As DM mass approaches the $W^\pm,Z$ masses, the $S_1S_1 \to VV$ ($V=W^\pm,Z$) annihilation channels open up and reduce the DM number density, which leads to an under production of DM for $m_{DM}$ above this range, irrespective of the CP-violating angle.

\begin{figure}[hhh!]
\begin{center}
\includegraphics[scale=0.55]{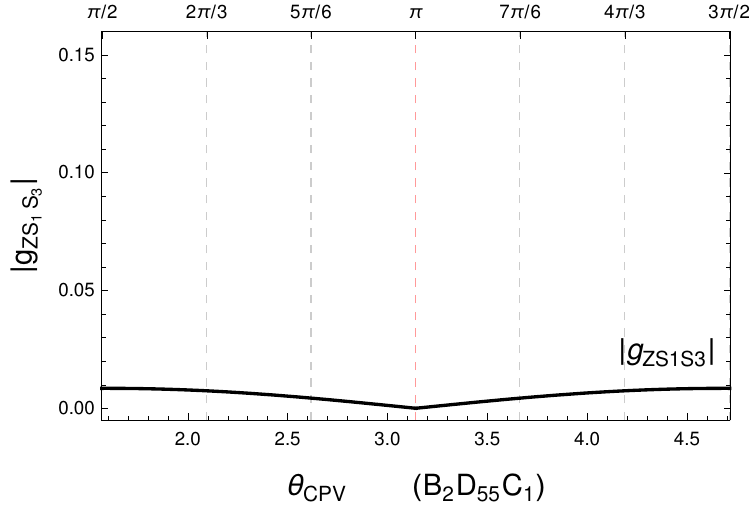}
\hspace{-2mm}
\includegraphics[scale=0.55]{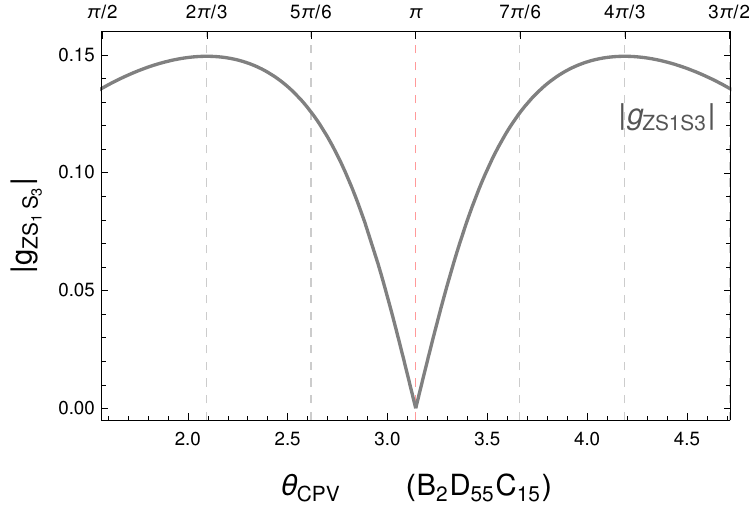}
\caption{The change of the $g_{ZS_1S_3}$ coupling with respect to the CP-violating angle in the type 2 scenarios for a given mass of $m_{S_1}=50$ GeV.}
\label{B2D55-gZS1S3-fig}
\end{center}
\end{figure}

\begin{figure}[ht!]
\begin{center}
\includegraphics[scale=0.55]{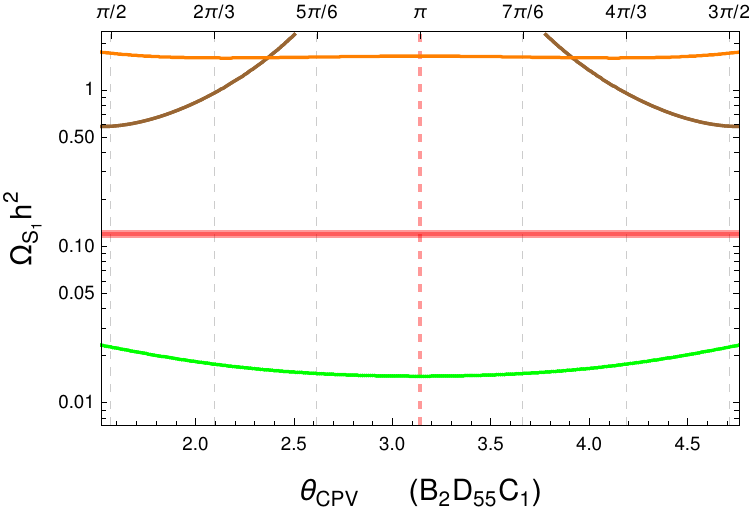}
\hspace{-2mm}
\includegraphics[scale=0.55]{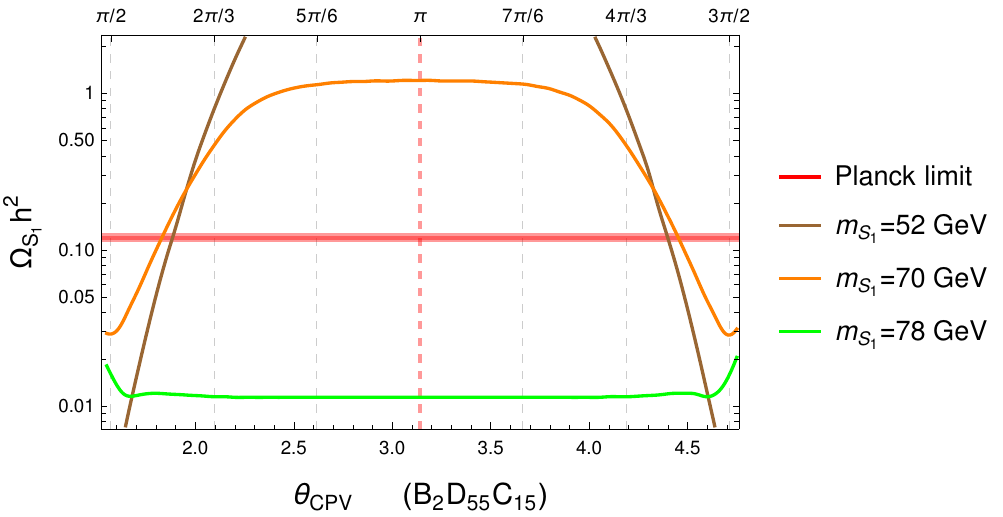}
\caption{The change in DM relic density for various DM masses with respect to the CP-violating angle. The horizontal red band shows the Planck observation limit on the abundance of DM.}
\label{B2D55-RelicAngle-fig}
\end{center}
\end{figure}

Recall that the $ZS_1S_3$ coupling is sensitive to the changes in the CP-violating angle. Figure \ref{B2D55-gZS1S3-fig} shows the absolute value of the $ZS_1S_3$ coupling for an exemplary DM mass of 50 GeV with respect to $\theta_{CPV}$ in type 2 benchmark scenarios.
Due to the striking difference between the values of the coupling in the two benchmark scenarios, one expects a substantial difference in the DM relic density in the two models. The B$_2$D$_{55}$C$_1$ scenario consistently over-produces DM in the mass range $45 \gev < m_{DM} < 75 \gev$ (except for large CP-violating angles around the $Z$ resonance region $m_{DM} \approx m_Z/2$ where $S_1$ and $S_3$ are very close in mass). In this scenario, the $ZS_1S_3$ coupling is so weak that it fails to co-annihilate DM effectively in this mass range. 
The B$_2$D$_{55}$C$_{15}$ scenario on the other hand, has a large enough $ZS_1S_3$ coupling at large $\theta_{CPV}$ to satisfy the Planck limit on the DM relic density.
Figure \ref{B2D55-RelicAngle-fig} confirms this behaviour where DM relic density for various DM masses is shown. The B$_2$D$_{55}$C$_1$ scenario over-produces DM for masses below 75 GeV, and the B$_2$D$_{55}$C$_{15}$ scenario produces DM in agreement with the Planck limit for large CP-violating angles for this mass range. 
As mentioned before, both scenarios under-produce DM for larger masses when the $S_1S_1 \to VV$ annihilation channel is open.

\begin{figure}[ht!]
\begin{center}
\includegraphics[scale=0.54]{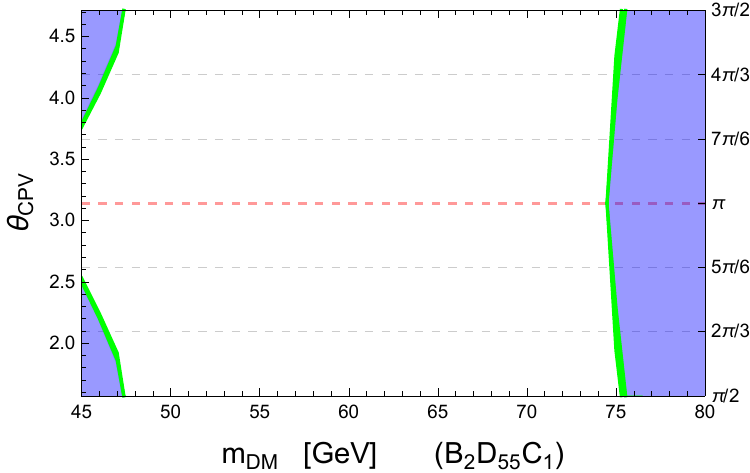}
\hspace{-2mm}
\includegraphics[scale=0.54]{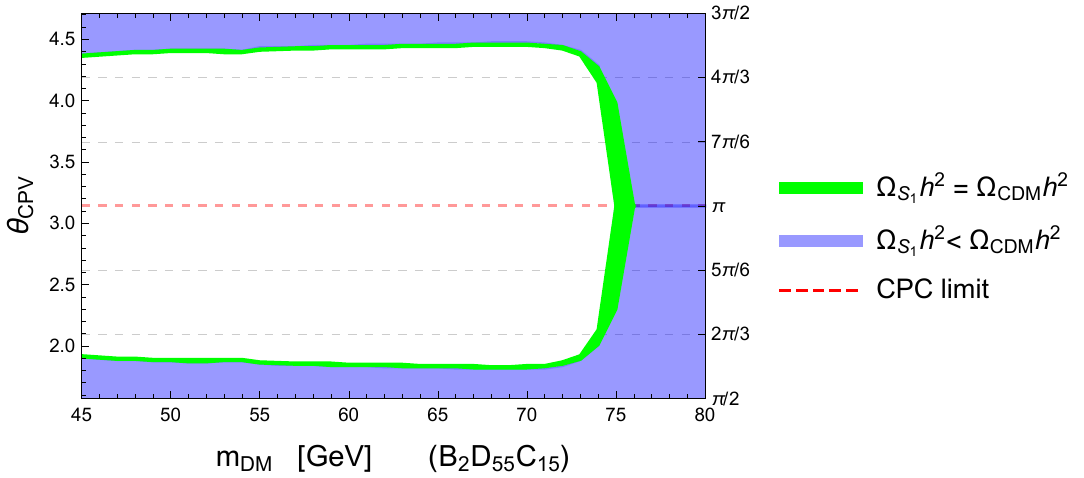}
\caption{Regions producing 100\% of DM in green and under-producing DM in blue in the $\theta_{CPV}$-$m_{DM}$ plane. The horizontal dashed red line represents the CP-conserving limit.}
\label{B2D55-AngleMass-fig}
\end{center}
\end{figure}

Figure \ref{B2D55-AngleMass-fig} shows regions where $S_1$ contributes to 100\% of the observed DM in green, and regions where it only provides a fraction of the observed relic density in blue, in the $\theta_{CPV}$-$m_{DM}$ plane. The blank regions are ruled out by Planck observations as they lead to an over-production of DM. Note that in the CP-conserving limit where $\theta_{CPV}=\pi$ the model fails to comply with the Planck observations. 

\subsection{Benchmarks of type 3}
The type 3 benchmark scenarios are defined in the heavy mass region, $m_{DM} \geqslant 80 \gev$, where the charged inert scalars could also be close in mass with $S_1$, thereby providing new co-annihilation channels for the DM candidate.
When studying the DM phenomenology of the model, it is not only the annihilation and co-annihilation of DM, but also the (co)annihilation of other inert particles amongst each other that should be taken into account.

\begin{figure}[ht!]
\begin{center}
\includegraphics[scale=0.56]{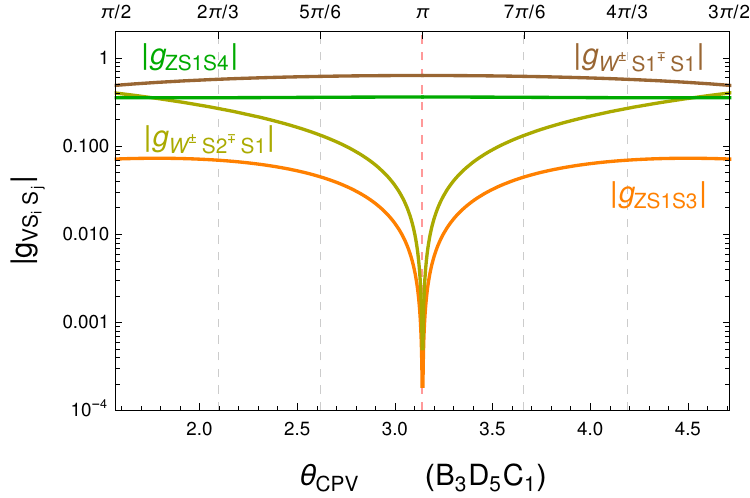}\\[2mm]
\includegraphics[scale=0.56]{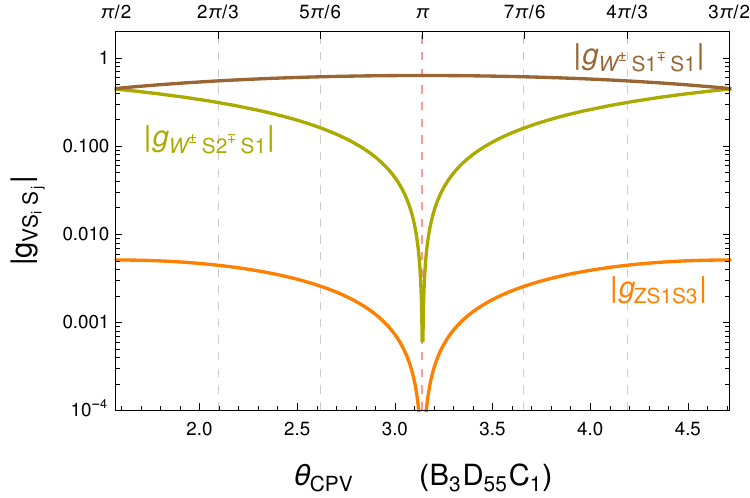}
\hspace{-2mm}
\includegraphics[scale=0.56]{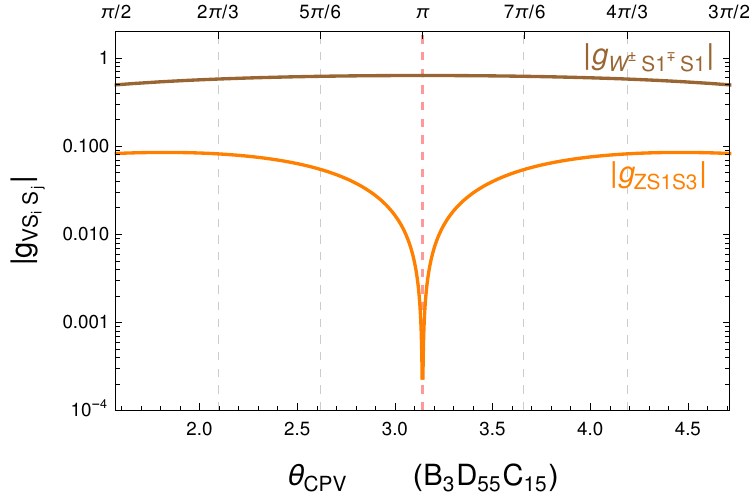}
\caption{The relative strength of the $g_{VS_iS_j}$ coupling in the two scenarios for a given mass of $m_{S_1}=90$ GeV.}
\label{B3D55-gVSiSj-fig}
\end{center}
\end{figure}

Figure \ref{B3D55-gVSiSj-fig} shows the relevant and non-negligible $VS_1S_i$ couplings for all three type 3 benchmark scenarios where $V=W^\pm,Z$ and $S_{i}$ is a neutral or charged inert particle. Due to the presence of so many co-annihilation processes, type 3 scenarios under-produce DM.
Note, however, that in the heavy mass region the annihilation $S_1S_1 \to VV$ is dominant whose coupling is independent of the CP-violating angle. Therefore, one expects a similar behaviour in all three scenarios.

\begin{figure}[ht!]
\begin{center}
\includegraphics[scale=0.6]{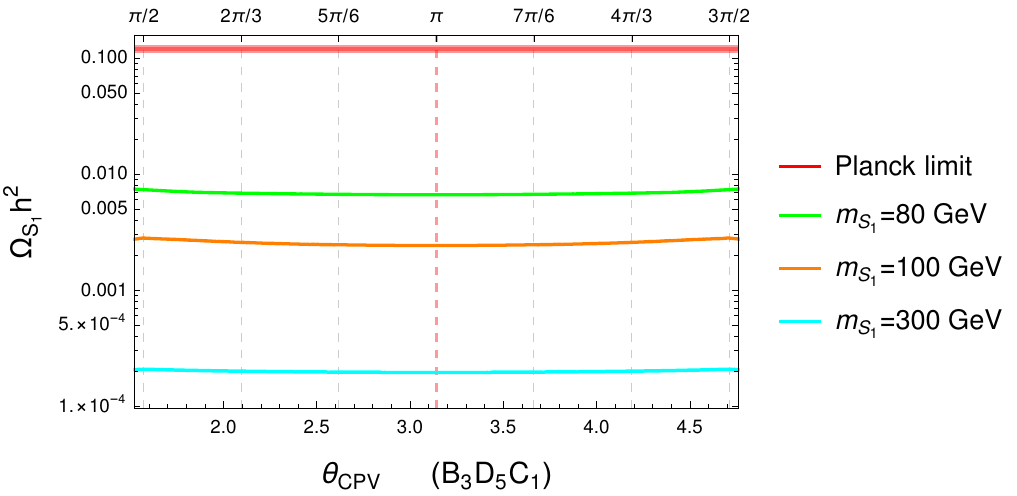}\\[2mm]
\includegraphics[scale=0.6]{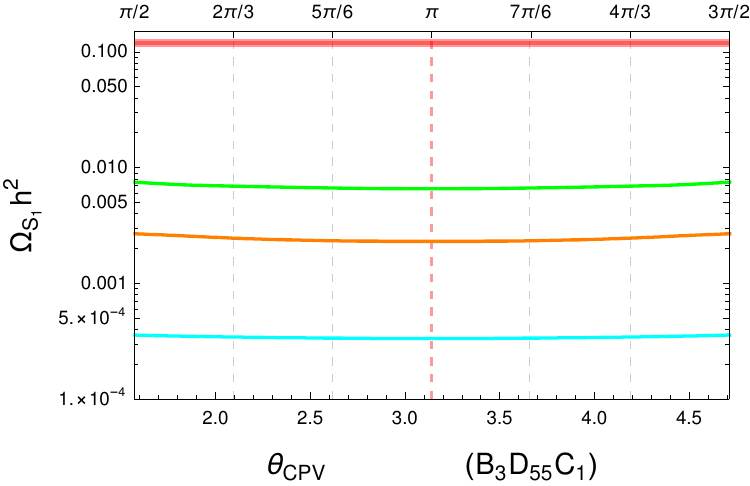}
\hspace{-2mm}
\includegraphics[scale=0.6]{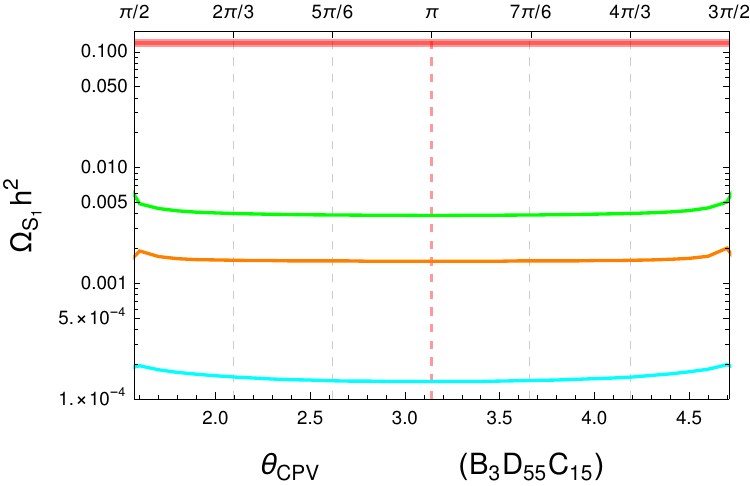}
\caption{The change DM relic density for various DM masses with respect to the CP-violating angle. The horizontal red band shows the Planck observation limit on the abundance of DM.}
\label{B3D5-B3D55-RelicAngle-fig}
\end{center}
\end{figure}

Figure \ref{B3D5-B3D55-RelicAngle-fig} shows the relic abundance of the DM candidate for various DM masses. As mentioned before, all three scenarios have a similar behaviour, with the B$_3$D$_{55}$C$_1$ scenario providing a slightly larger relic density in comparison to the B$_3$D$_{5}$C$_1$ scenario due to a larger $\delta{12}$ mass splitting.
Scenario B$_3$D$_{55}$C$_{15}$ provides only two co-annihilation channels for $S_1$; however, they have larger couplings compared to the B$_3$D$_{55}$C$_1$ case, which leads to a slightly smaller relic abundance for $S_1$.
The reason we do not see the revival of the very heavy mass region $m_{DM} > 400$ GeV as shown in \cite{Cordero-Cid:2016krd,Keus:2015xya} is the absence of the Higgs mediated processes 
$S_i S_j \to h \to VV$, where $S_{i,j}$ is any neutral or charged inert particle. These Higgs-mediated processes have a destructive interference with pure gauge processes $S_i S_j \to VV$, which would have revived the heavy mass region.

\section{Conclusion and outlook}
\label{conclusion}

The scalar potential is the least constrained sector of the SM which, if extended, could provide new sources of CP-violation and viable DM candidates. 
New sources of CP-violation are often limited due to their contribution to the EDMs. Introducing CP-violation in the dark sector, however, is unconstrained as it has no contribution to the EDMs.
On the other hand, dark sectors interacting with the visible sector through a Higgs portal are under tension, as they require a large portal coupling for efficient annihilation of DM and a small portal coupling to satisfy direct and indirect detection experiments and SM-Higgs data.

We present a novel mechanism in which the CP-violating dark particles only interact with the SM through the gauge bosons, primarily the $Z$ boson, in the framework of a three Higgs doublet model. 
In the region where Higgs portal interactions are sub-dominant, we show that the $Z$ portal CP-violating DM can still thermalise and satisfy all experimental and observational data. 

In the context of electroweak baryogenesis, the extended scalar sector could easily accommodate a strong first order phase transition. The efficient transfer of the unconstrained dark CP-violation to the visible sector is under study and will be the subject of our future publication.

\subsection*{Acknowledgement}
The author would like to thank D.~Sokolowska for useful discussions,
and D.~Weir, K.~Rummukainen and K.~Tuominen for their invaluable support in an environment which is not always conducive to gender equality. 
The author acknowledges financial support from H2020-MSCA-RISE-2014 Grant No.~645722 (NonMinimalHiggs), the Research Funds of the University of Helsinki, Academy of Finland projects ``Particle cosmology and gravitational waves" 
No.~320123 and ``Particle cosmology beyond the Standard Model" No.~310130.

\end{document}